# On Topological Structure of Web Services Networks for Composition

## Chantal Cherifi* and Jean-François Santucci

SPE Laboratory, Corsica University, Corte, France
E-mail: chantalbonner@gmail.com
*Corresponding author

**Abstract:** In order to deal efficiently with the exponential growth of the Web services landscape in composition life cycle activities, it is necessary to have a clear view of its main features. As for many situations where there is a lot of interacting entities, the complex networks paradigm is an appropriate approach to analyze the interactions between the multitudes of Web services. In this paper, we present and investigate the main interactions between semantic Web services models from the complex network perspective. Results show that both parameter and operation networks exhibit the main characteristics of typical real-world complex networks such as the "small-world" property and an inhomogeneous degree distribution. These results yield valuable insight in order to develop composition search algorithms, to deal with security threat in the composition process and on the phenomena which characterize its evolution.

**Keywords:** Web services, Composition, Interaction networks, Complex networks, Semantics

## 1. Introduction

From a simple documentary system, the Web grew to become a huge network of distributed data, applications and users where people interact. However, this ever growing amount of information is a major challenge to achieve an efficient use of Web technologies. In order to render Web content meaningful both for human and machine consumption, the semantic Web has developed a set of ontology-based technologies, tools and standards. Building on this framework, Web services provide a rapid way to share and distribute information between clients, providers and commercial partners. These modular applications, independent of any software or hardware platform, can be coupled through the Web to create new value-added services. This composition process is one of the most challenging aspects of Web services. Indeed, it must face the issues of heterogeneity, volatility and growth to provide solutions ensuring high availability, reliability and scalability. The development of Web services thus raises similar difficulties to those that accompanied the growth of the Web.

In their current state, Web services suffer from interoperability necessary in order to link exchanged data. In order to address this requirement, the semantic Web introduced semantics in Web services, on the top of the basic tools of Web services (SOAP, WSDL, UDDI). By enriching the descriptions with semantic concepts, it intends to allow the automation of publication, discovery and composition.

However, the benefit of semantic Web services is hindered by the ever growing amount of information provided by this living complex system. Besides this growing aspect, the Web services space is highly dynamic as services are susceptible to changes, relocations and suppressions. Understanding the characteristics that hold together the Web services in this overall complex system is of prime interest. It should lead to more efficient solutions of Web services composition lifecycle management.

In order to tackle complexity, a network based paradigm has recently emerged. The basis of this approach relies on the simplicity of the model to predict the behaviour of a system as a whole from the interactions of its constituents. The system is described using graph theory where individuals are the nodes of the graphs and a link between two nodes exists if they interact. The so-called network science has built tools to address the inner workings of complex systems of various origins. It has been very successful to solve the difficulties that many research fields face from biology to computer science (Newman, Barabasi, and Watts, 2011; Scott, 2000).

The investigation of real-world systems, characterized by a decentralized and apparently unplanned evolution, showed that, irrespective of their origin, they share a common structure. They typically exhibit the "small-world" property, i.e. any node in the network can be reached, on average, passing through a small number of nodes. Besides, the distribution of the number of links attached to the nodes is highly heterogeneous. The great majority of nodes have few interactions, while few nodes are highly connected. Generally, the node degree distribution can be adequately modelled by a power law. In this case the network is said "scale-free". Furthermore, a community structure is observed in most real-world networks, i.e. nodes are organized into highly cohesive sets called modules or communities.

One point of paramount importance is that these common features allow developing common tools to understand and to process these networks. There has been some important advances, particularly on network resilience (failures and attacks in technological networks), epidemiological processes (spread of diseases in social networks, viruses in computer networks) and search on networks (Web crawling, files on distributed databases).

Similarly to other complex systems, the set of interacting Web services can be represented by a graph in order to represent the Web service composition structure. According to node and link definitions, different network models can be defined. In Web services networks, nodes are Web services and a link account for an interaction between any operations of two services. In operation networks nodes are operations and a link is drawn between two operations if one can furnish the information necessary to invoke the other. Finally, in parameter networks, nodes are parameters and a link represents an input/output relationship between two parameters.

Although very promising, few authors have proposed to use the complex network paradigm in the Web services context. In (Kil et al., 2009), the authors investigate syntactic Web services networks topology. Experimental results show that these networks exhibit the small-world and scale-free properties. Based on these results, they define a generator producing synthetic benchmarks of WSDL description files. Web services networks have also been used to address the composition search problem. Usually, the method to find a set of interacting Web services that satisfy a given request consists in a network mining. In (Talantikite, Aissani, and Boudjlida, 2009), compositions are discovered within a semantic Web services network by a forward chaining algorithm. In (Liu and Chao, 2007), search of compositions in a syntactic Web services network is performed using graph matching techniques. In (Kwon et al., 2007), a semantic Web services network is stored in a relational database. Composition search is done by SQL statements. In (Hashemian and Mavaddat, 2005), a breadth first search algorithm is used to search for compositions in a semantic network of parameters. Note that in all these works, no information on the topological properties of the network is used. To our knowledge, the only attempt to exploit a topological property of the network in order to design a composition algorithm is related in (Gekas and Fasli, 2007). The authors propose a composition algorithm guided by the link analysis of a semantic Web services interaction network. The underlying idea is that the relevance of a service to enter into a composition is related to its importance in terms of connectivity with its neighbourhood. One can thus associate a rank to each node reflecting its reputation. This information is used by the A* shortest path search algorithm to probe the Web services space.

Although networks are an appropriate representation to deal with Web service composition, there is a crucial need to better characterize the topological properties of these networks. Indeed, such knowledge can be exploited in various ways. For example, efficient composition search strategies can be developed. Web services security and dynamic problematic can be also addressed in a new light.

Our goal in this work is to investigate the structure of the Web services space under the complex network framework. This article goes beyond the state of the art through the main following contributions:

1. We extend the topological analysis of composition networks to semantic networks. Besides the basic topological properties, we consider assortativity and community structure.

2. To get a more detailed analysis of interactions, we consider the operation granularity rather than the Web service one. Using ontological subsumption relationships, we define four different operation semantic network models that reflect more or less effective compositions. This approach differs from previous work on network models where semantic relationships are not clearly defined or merged in a single dimension.

3. We perform an extensive comparative evaluation of the topological properties of the operation and parameter networks. We discuss the influence of the identified properties on the composition process. The impact of the networks structure on dynamic processes, particularly nodes failure, is analysed. We give some guidelines on the networks topological features that can be used to optimize a composition search in the networks.

The rest of the paper is organized as follows. Web services and their description are introduced in section 2. In section 3 we describe operation and parameter network models design principles. Section 4 reviews the complex network topological properties under investigation. Section 5 is devoted to the experimental results. We present the Web service collection used to build the Web services networks. A comparative analysis of the topological properties is given. We provide some guidance on how to take advantage of those properties in order to tackle the composition search problematic. Conclusion and future work are presented in section 6.

**2. Semantic Web services**

The Web services composition life cycle includes publication, discovery, composition synthesis, orchestration, and control during execution process. The description of Web services capabilities is a key element for Web service discovery, composition and management. It is used to guide service discovery and selection and to determine their interaction.

Currently, Web services description is based on the Web Service Description Language (WSDL) standard. WSDL is written in XML which specifies functional and non-functional information. Functional information embodies operations that a service exposes with their input and output parameters. Non-functional information concerns the location of the service, the communication protocol and the data format specifications. Since WSDL provides only syntactical information, the semantics implied by the provider can lead to possible misinterpretation by others. Adding semantics to Web service descriptions can be achieved by using ontologies that support shared vocabularies and domain models.

The semantic Web service field includes substantial bodies of work trough three conceptual approaches, Ontology Web Language for Services (OWL-S) (Martin et al., 2004), the Web Services Modelling Ontology (WSMO) (Lausen et al., (eds) 2005) and Semantic Annotation for WSDL (SAWSDL) (Farrell and Lausen, 2007). OWL-S and WSMO adopt a top-down perspective. The semantics is described independently of the service development. It is grounded with the corresponding WSDL description. SAWSDL adopt a bottom-up approach where the WSDL file is enriched with semantic information.

OWL-S is a specific ontology for Web services written in the Ontology Web Language (OWL). This high-level ontology includes three upper ontologies: profile, process model and grounding. The profile specifies inputs and outputs, preconditions and effects, all linked to ontological concepts. It is devoted to the functional aspects of Web services. The process model provides a runtime framework to monitor service execution. The grounding gives information on how to use the service. It provides a pragmatic binding between the profile, the model and the physical Web service location.

WSMO design principles are expressed through four top-level elements. Goals represent the client's objectives. Web service descriptions describe the functional and behavioural aspects of Web services. Ontologies provide the terminology used by other WSMO components. Mediators are connectors between components to provide interoperability between ontologies. Web Service Modelling Language (WSML) is WSMO's own service description language to represent functional and non-functional semantics of Web services. Like OWL-S, it is a structured and logic-based language.

The SAWSDL standard semantic Web service description language allows for a structured representation of service semantics in XML, with references to any kind of non-logic-based or logic-based ontology for semantic annotation. It establishes mapping between existing WSDL elements and ontological concepts. The "model reference" extension allows specifying associations between WSDL components (interfaces, operations and their input and output messages) and an ontological concept. Data mediation is performed at ontology level through lifting and lowering schema mapping. Lifting schema mapping is used to convert data XML schemas to ontology schemas. Lowering schema mapping converts ontology schemas into data XML schemas.

Despite its undoubted strengths, the introduction of semantics in Web services description raises new problems. The data semantics may be expressed by different domain ontologies. Mapping could provide a common layer from which several ontologies could be accessed and hence could exchange information in semantically sound manners. This problem is a self-contained field that has triggered a large amount of works (Shvaiko and Euzénat, 2005). This promising area for the management of diversity and heterogeneity of distributed information sources is not considered in this work. In the following, two concepts that do not originate from the same ontology cannot be compared and consequently cannot be similar.

In this paper, we focus on the functional aspect of Web services. Hence, we restrict the definition of a Web Service to a set of operations with their parameters. We use the following notations. A Web service is a set of operations. Its name is represented by a Greek letter. Each operation numbered by a digit contains a set of input parameters noted I and a set of output parameters noted O. Each parameter described by an ontological concept is represented by a lowercase letter. Figure 1 represents a Web service α with two operations 1 et 2, input parameters $I_1 = \{a,b\}$, $I_2 = \{c\}$, output parameters $O_1 = \{d\}$, $O_2 = \{e,f\}$. In the following we can use for short the word parameter rather than "parameter concept".

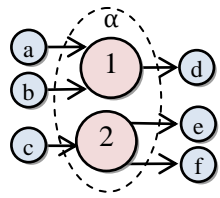

Figure 1. Schematic representation of a Web service α with two operations 1 and 2.

**3. Web service network models**

A Web service network represents interacting entities that can be parameters, operations or more generally Web services. Although at first glance using services as nodes rather than operations seems more natural, it is more appropriate to consider the latter. Indeed, operations are the main point of interest when it comes to Web services composition; they are the atomic interacting entities. In the following, we will therefore restrict our attention to networks defined with either parameters or operations as nodes.

In a parameter network, a node is a parameter concept and a link represents the dependency relation between an input and an output parameter of an operation; it materializes an operation. In an operation network, a node is an operation and a link represents an elementary composition between two operations; it materializes the fact that an operation can provide the data necessary to invoke another one. Although these networks are very different in nature, we call them interaction networks. Indeed, they reflect in different ways the interaction relationships between a set of Web services in a composition process.

*3.1 Interaction network of operations*

An interaction network of operations is a directed graph where nodes represent the Web services operations and relationships materialize an information flow between operations. Let i be a source operation described by its sets of input and output parameters ($I_i$, $O_i$). Let j be a target operation described by ($I_j$, $O_j$). There exists an interaction relationship between i and j if i is able to invoke j, that is, if and only if for each input parameter of j, there is a similar output parameter of i. Service compatibility is therefore reduced to semantic parameter matching. Many diverse solutions dealing with the similarity between concepts have been proposed so far. Two main families of approaches can be identified for the calculation of such ontology-based similarity measures: those based on shared information (Hau and Lee, 2005; Resnik, 1995; Couto and Silva, 2011) and those using only hierarchical relationships (Paolucci et al., 2002; Rada et al., 1989).

To achieve the comparison, we take as a basis the classical subsumption relationships introduced in (Paolucci et al., 2002). Defined for service discovery purpose, it is expressed by four degree of match: *exact*, *plugin*, *subsume* and *fail*. We introduce a fifth degree named *fitin* in order to broaden the service compatibility. Note that any other semantic matching similarity function based either on the ontology structure or on the shared informative content of the concepts can be used to define new interaction networks.

Let two concepts to be compared, $C_o$ (output concept) the concept of the invoking operation and $C_i$ (input concept) the concept of the invoked operation.

In an *exact* matching, $C_o$ is similar to $C_i$ if they are described by the same ontological concept. Let consider the ontology fragment of Figure 2 and two operations. The first operation takes as input a school level ($C_i$ = `SchoolLevel`) and provides a biology textbook ($C_o$ = `BiologyTextbook`). The second operation takes as input a biology textbook ($C_i$ = `BiologyTextbook`) and provides the corresponding price ($C_o$ = `Price`). The first operation can invoke the second one because the

`BiologyTextbook` concepts are equivalent. The composition allows to obtain the price of biology textbooks for a given school level. As a result, the goal of the invoking operation is fully satisfied and the full capabilities of the invoked operation are used.

In a *plugin* matching, $C_o$ is similar to $C_i$ if $C_o$ is more specific than $C_i$. Let consider the previous first operation which takes as input a secondary school level ($C_i$ = `SchoolLevel`) and provides a list of biology textbooks ($C_o$ = `BiologyTextbook`). The second operation takes as input all types of textbooks ($C_i$ = `Textbook`) and provides the corresponding price ($C_o$ = `Price`). In this case, the first operation can still invoke the second one because the concept `BiologyTextbook` is more specific than the concept `Textbook`. Similarly to the previous case, the composition allows to obtain the price of biology textbooks for a given school level. However, only part of the capabilities of the invoked operation is used. Indeed, in order to obtain the price of another type of textbooks, one needs to change the composition or add some more invoking operations.

In a *subsume* matching, $C_o$ is similar to $C_i$ if $C_o$ is more general than $C_i$. Let consider the same first operation which takes as input a secondary school level ($C_i$ = `SchoolLevel`) and provides a list of biology textbooks ($C_o$ = `BiologyTextbook`). Now, the second operation takes as input an anatomy textbook ($C_i$ = `AnatomyTextbook`) and provides its price ($C_o$ = `Price`). The first operation can invoke the second because the `BiologyTextbook` concept is more general than the `AnatomyTextbook` concept. However, in this case, the composition can provide the price of the anatomy textbooks for a given school level rather than the price of all biology textbooks. The goal of the composition to obtain the price of the biology textbooks is in this case only partially satisfied. To get the prices of all type of biology textbooks, one should consider the invocation of additional operations.

In a *fitin* matching, $C_o$ is similar to $C_i$ either if they are described by the same ontological concept, or if $C_o$ is more specific than $C_i$. In other words, the fitin operator embodies the exact and the plugin matching. In both cases, the composition goal is always fully satisfied.

The *fail* matching means that there is no subsumption relation between $C_o$ and $C_i$. No interaction is possible between the two considered operations.

The matching degrees can be ranked according to their relevance as followed: *exact > fitin > plugin > subsume > fail*. An exact match will be the best possible match between an output parameter of an operation and an input parameter of another operation. The fitin match is better than the plugin one because it allows for both plugin and exact relationships within the same interaction. Finally, the subsume match allows recovering less relevant interactions than the previous operators.

Starting from the concept similarity, we can now define an interaction relationship from an invoking operation i to an invoked operation j. For the sake of clarity, we choose to define a network for each degree of match where the nodes are the operations and a link between two nodes is drawn if and only if all the parameters share the same degree of match. In the case of the fitin relationship, three situations may occur: all the parameters have an exact degree of match, all the parameters have a plugin degree of match, some of them have an exact degree of match and the others have a plugin degree of match. Each of these full invocation networks provide a particular and a complementary view of the relationships that can exist within a set of operations.

Note that one can use a less restrictive definition to draw a link between two operations. Indeed, one can build a composition even if just a subset of the input parameters needed to invoke an operation is provided. This partial invocation gives rise to a correct composition if the non-provided parameters are optional. Otherwise, it involves using additional operations to fully fulfil a composition goal. Partial invocation allows more composition possibilities, but it is less effective than full invocation.

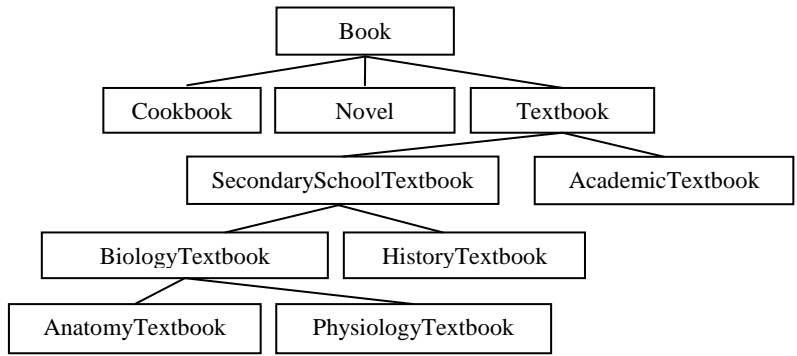

Figure 2. Fragment of a book ontology.

Figure 3(b) represents an exact operation network extracted from four operations numbered from 1 to 4, belonging to three Web services α, β and γ in Figure 3(a). Their input and output parameter concepts are labelled from a to i. Operation 2 can invoke operation 3 because their respective output and input parameter f have an exact degree of match. Furthermore, all the entries of operation 3, $I_3 = \{f\}$, are included in the output of operation 2, $O_2 = \{e, f\}$. For this reason, there is a link from operation 2 to operation 3 in the interaction network. For the same reasons, there is a link between operations 3 and 4.

### 3.2 Interaction network of parameters

An interaction parameter network is defined as a directed graph in which nodes represent the set of parameter concepts and links materialize operations. Each operation i can be defined as a triplet $(I_i, O_i, K_i)$, where $I_i$ is the set of input parameters, $O_i$ is the set of output parameters and $K_i$ is the set of parameters dependencies. To build the set of interdependencies, we consider that each output parameter of an operation depends on each input parameter of the same operation. In other words, a link is created between each concept associated to an input parameter of an operation and each concept of its output parameters. Connectivity within a parameter network is partly due to the fact that some parameter concepts can be used by several operations. Moreover, they can be used as input parameters by some operations and as output parameters by other. In the network, similar parameters are represented by the same node. To assess the similarity between two parameter concepts, we consider *exact* and *fail* degrees of match defined previously. Other matching degrees cannot be used in this case because it would lead to inconsistent groupings. Figure 3(c) represents a parameter network build from the four operations of Figure 3(a), and their nine input and output parameter concepts. To illustrate the parameters dependencies, consider for example {f, g, h} concepts. They appear more than once, either as input or as output of several operations. f is an output of operation 2 and an input of 3, g and h are outputs of 3 and inputs of 4. These parameters are respectively represented by a single node in the network.

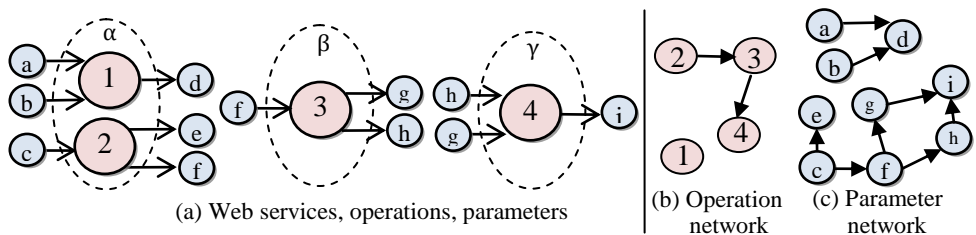

(a) Web services, operations, parameters   (b) Operation network   (c) Parameter network

Figure 3. Interaction network of operations with 4 nodes (b) and interaction network of parameters with 9 nodes (c) obtained from four operations (a).

## 4. Complex networks

Systems formed with a large number of interacting individual elements can be adequately described by a network where individuals are the nodes of the network and a link between two nodes is drawn if they interact. Neural networks, metabolic networks, the Internet, the World Wide Web, social networks, are typical examples of such systems. Many concepts and statistical measures have been designed to capture in quantitative terms their underlying organizing principles. Analysis results has led to the conclusion that despite their many differences, such complex networks are governed by common laws that determine their behaviour (Costa et al., 2007). As our first goal is to check if Web services based networks share these properties, we concentrate on measurements of the overall structure rather than on local properties of nodes. In the following, we recall the definition of the most useful concepts which summarize the essential of a complex network structure.

*4.1 Small-world*

In "*small-world*" networks, most nodes are not neighbours of one another, but most nodes can be reached from every other by a small number of links. This property was demonstrated by Milgram's experimental study on the structure of networks of social acquaintance. Results showed that a chain of "A friend of a friend" can be made, on average, to connect any two people in six steps or fewer. This experience gave rise to a myth that became popular under the statement of "Six degrees of separation". Small-world is therefore a notion related to the network distance between two nodes, which is defined as the number of links in the shortest path connecting them. In small-world networks, the average distance over all pairs of nodes is low and it varies with the total number of nodes, typically as a logarithm (Newman, 2003). The existence of shortcuts connecting different areas of the network can be interpreted as propagation efficiency. The small-world property has been observed in a variety of real-world networks. For example, the Web network where pages are nodes and links represent hyperlinks between pages, is a small-world network. Its average distance value is 18.59 for $8.10^8$ nodes (Albert, Jeong, and Barabasi, 1999). This phenomenon even occurs in the random Erdős-Rényi networks where each pair of nodes is joined by a link with probability p at random. Comparing the average distance of some networks of interest to the one estimated for Erdős-Rényi networks containing the same numbers of nodes and links, allow assessing their small-world property.

*4.2 Clustering*

Clustering, also known as transitivity, quantifies how well connected are the neighbours of a node. It is a typical property of friendship networks, where two individuals with a common friend are likely to be friends. A triangle being a set of three vertices connected to each other, the clustering is formally defined as the triangle density of the network. It is obtained by the ratio of existing to possible triangles in the considered network (Newman, 2003). Its value ranges from 0 (the network does not contain any triangle) to 1 (each link in the network is a part of a triangle). As a difference with the classical Erdős-Rényi random graph model, social networks are typically characterized by a high clustering coefficient. Others, such as technological and information networks exhibit a low transitivity value (Boccaletti et al., 2006). Following their work on the US power grid, the actor collaboration network and the neural network of Caenorhabditis elegans, the authors of (Watts and Strogatz, 1998) found that the small-world property is stressed by the proportion of triangles in a network. Indeed, the fact that two neighbours of a node are themselves neighbours, contribute to reduce the distance in a network.

*4.3 Assortativity*

Assortativity allows to qualify how nodes tend to associate together. It expresses possibly existing preferential attachments between them. For example, in social networks, people tend to connect to each other according to some shared features. They may tend to associate preferentially with people who are similar to themselves in some way. That is what we call *assortative* mixing. The number of links connected to a node, referred as the node degree, is the most prominent similarity criterion used. It can be interpreted as a measure of the leadership of a node in the network. In this case, the degree correlation reveals the way nodes are related to their neighbours according to their degree. A network is said to exhibit assortative mixing if nodes are preferentially linked to others with similar degree. It is called disassortative otherwise. This property is measured by the degree correlation (Boccaletti et al., 2006). It takes its value between -1 (perfectly disassortative) and 1 (perfectly assortative). Social networks generally tend to be assortatively mixed, while other kinds of networks are generally disassortatively mixed (Newman, 2003).

*4.4 Degree distribution*

The degree distribution has significant consequences for our understanding of natural and man-made phenomena as it is particularly revealing of a network structure. Typically, random networks are "homogeneous". Their nodes degree tends to be concentrated around a typical value. In contrast, many real-world networks are highly inhomogeneous with a few highly connected nodes and a large majority of nodes with low degree. Such networks tend to have quite heavy tailed degree distribution, often described by a power-law of the form $p_k \approx ck^{-\gamma}$, with values of $\gamma$ typically between 2 and 4. The so-called *scale-free* networks emerge in the context of a growing network where new nodes connect preferentially with existing nodes with a probability proportional to their degree. This preferential attachment is illustrated by the expression "The rich get richer". Networks can be characterized by different inhomogeneous distribution such as truncated *scale-free* networks, characterized by a power-law connectivity distribution followed by a sharp cut-off with an exponential tail. Note that for directed networks, three degree distributions can be estimated: out-degree distribution for outgoing links, in-degree distribution for ingoing links, and joint in-degree and out-degree distribution (Costa et al., 2007).

*4.5 Community structure*

In social systems, individuals gather within communities that represent the fundamental level of a society organisation. Many other systems of interest from various origins are found to naturally divide into communities. In complex networks, the presence of communities is revealed by the presence of fairly independent groups of nodes, with a high density of links between nodes of the same group and a comparatively low density of links between nodes of different groups. The community structure is a network feature that depicts the organisation of nodes into communities. Many community detection algorithms have been proposed in order to discover the community structure (Fortunato, 2010). They partition the network into a set of overlapping or non-overlapping communities. The community structure can be characterized by the number of communities, the community size distribution and the modularity.

The *modularity* is a global measure of the communities cohesiveness (Newman and Girvan, 2004). It compares the actual proportion of community internal edges to the expected edges proportion if links are randomly distributed. Its value ranges from -1 to 1. For networks exhibiting no community structure or when communities are no better than a random partition, the modularity value is negative or equal to 0. In the case of a community structure, the modularity value is between 0 and 1. Practically, a value between 0.3 and 0.7 is considered to be high (Newman, 2006). The modularity is very often used as a reference measure to evaluate the quality of a network partitioning.

The *community size distribution* is an important feature of a community structure. The studies conducted so far on real-world networks tend to show that the community size distribution follows a power-law (Newman, 2004; Guimerà et al., 2003) with an exponent between 1 and 2. In other words, community size is heterogeneous with the presence of a few large communities and many small ones.

*4.6 Organisation by components*

Real-world complex networks are generally divided into independent sub-networks called components. The size of the largest one is an important quantity. For example, in a communication network like the Internet, the size of the largest component represents the largest fraction of the network within which communication is possible. It is hence a measure of the effectiveness of the network at doing its job (Newman, 2003). A component is a (maximal) subset of vertices connected by paths through the network. Studies on the organisation of components generally focus on the components size in order to express the network's global topology.

**5. Analysis of Web service networks**

In this section, we investigate the topological properties of operation and parameter networks. First of all, we present the Web services benchmark used in our experiments to extract the networks. Four operation networks corresponding to the four level of similarity defined between parameters (exact, plugin, subsume, fitin), and a parameter network have been build using these data. We then perform a comparative evaluation of the operation networks in order to evaluate the impact of the similarity function. Starting from the overall organisation, we then focus on the largest component to compute the main topological properties i.e. small-world, degree distribution, clustering, assortativity and community structure. Finally, we provide a comparison between the exact operation and parameter networks. Based on this comparison, we show how the network structure can be used for composition search, and which networks are the most appropriate for composition discovery and why, according to their features.

*5.1 Web services benchmark*

Different benchmarks of publically available Web services description collections are available. They are provided by different entities like the ICEBE organisation (IEEE International Conference on e-Business Engineering, 2005), the ASSAM WSDL Annotator project (Hess, Johnston, and Kushmerick, 2004), SemWebCentral (InfoEther and Technologies, 2004), OPOSSum (Küster, König-Ries, and Krug, 2008) or even the authors of (Fan and Kambhampati, 2005). Three of them gather semantic Web services descriptions. ASSAM Dataset2 and SemWebCentral SWS-TC contain respectively 164 and 214 OWL-S descriptions. As their size is rather limited, we did not retain these samples in our study. SemWebCentral provides the same set of services using three description languages (SAWSDL, OWL-S and WSDL). These collections called SAWSDL-TC1 (SAWSDL Test Collection Version n°1) and its counterpart OWL-TC1 are large enough to be considered. We choose to concentrate on SAWSDL-TC1 because it also includes the syntactic descriptions. In order to investigate the impact of semantics on network structures, syntactic and semantic representations have been compared in (Cherifi, Labatut, and Santucci, 2010b; Cherifi, Labatut, and Santucci, 2010a). Although it has been initially designed to evaluate Web services discovery algorithms, the collection is a representative sample of Web services that may interact within a composition. It has been re-sampled to increase its size and contains 894 descriptions among which 654 are classified into seven domains (*economy*, *education*, *travel*, *communication*, *food*, *medical* and *weapon*). Each description contains only one operation. The collection contains 2136 parameter instances that are semantically described by their ontological concept. Using this collection, we extracted a set of five networks with WS-NEXT, a network extractor toolkit specifically designed for this purpose (Cherifi, Rivierre, and Santucci, 2011). These networks (exact operation network, plugin operation network, subsume operation network, fitin operation network, exact parameter network), defined in section 3, have been used throughout the experiments.

*5.2. Structure and comparison of operation networks*

All the networks share the same global structure. A "giant" component stands along with a set of small components and isolated nodes. The proportion of these three elements is presented in Table 1. The number of nodes is the same in all the networks (785). It corresponds to the number of operations in the collection. Operations are globally equally dispatched between isolated nodes and the giant component while the small components contain a lower proportion of nodes. This structure reflects the decomposition of the collection into several non-interacting groups of operations. The fitin network is the one that attracts the highest percentage of operations in the giant component. It also contains the lower percentage of isolated nodes. This is due to a less restrictive matching function. Indeed, it is easier to create links with this matching function which includes two types of relationships (exact and plugin). Among the remaining networks, the plugin one has the highest percentage of isolated nodes. Besides, the percentage of nodes in the small components is the lowest. This shows that when there is a subsumption relationship between two concepts, the situations where the input concepts are more general than the output concepts are the less numerous. Accordingly, the subsume network has the lowest percentage of isolated nodes. Indeed, the matching function in this case is complementary to the plugin one. Note that the number of nodes in the small components of the subsume network is quite high. It represents one quarter of the giant component's nodes. We also report the proportion of nodes in both small and giant components. Indeed all the operations in these components can be composed. According to this criterion, the networks can be classified in the following order: fitin, subsume, exact, plugin. The effectiveness of the fitin network is tied to the matching function as reported earlier. The second place of the subsume network reflects the fact that, in this benchmark, Web services developers had a slight tendency to use ontological concepts associated with output parameters more general than the ones associated with the inputs.

Table 1. Proportion of nodes in different elements of the operation networks.

| Network | Isolated nodes | Small components | Giant component | Small + Giant components |
|---|---|---|---|---|
| plugin | 50.58% | 2.42% | 47.00% | 49,42% |
| exact | 48.79% | 7.77% | 43.44% | 51,21% |
| subsume | 45.99% | 12.10% | 41.91% | 54,01% |
| fitin | 42.00% | 6.50% | 51.50% | 58,00% |

*5.2.1 Isolated nodes*

The number of isolated nodes is pretty high in all the networks. They represent operations that cannot invoke or be invoked by other operations. None of their output parameters can serve as input and none of their input parameters is provided by other operations. Isolated nodes can only be invoked as atomic operations. As such, they do not represent any added value for composition. Providers should develop Web services gateways to connect them to dense areas of the giant component for this purpose. The content of the sets of isolated nodes is very similar in all the networks. Indeed, the overlapping rate of the sets of isolated nodes of different networks varies from 79% to 87%. As the proportion of small components is not very high, this suggests that the sets of nodes that belong to the giant component of the different networks are also very similar. The main difference lies in the way those nodes are linked.

*5.2.2 Small components*

The number of small components is very different according to the networks definition. The fitin, plugin, subsume and exact networks respectively contain 2, 3, 4 and 7 small components. To better understand their content and organisation, we performed a visual analysis. To do so, the small components of the exact network have been labelled from 1 to 7 and we followed their trace according to the network definition. Globally, small components have a star structure i.e. they are organized around a hub or an authority. Components numbered 1, 2 and 3 are made with operations originating from the travel domain while operations in components 4, 5, 6 and 7 are from the education domain. Component 1 is the only one that is present in all the networks. Components 2 and 3 merge in a single component in the fitin and subsume networks. Similarly component 4 and 5 merge in the subsume network. Component 3 exists also in the plugin network. The remaining components in the plugin and subsume networks are distinct components, either grouping nodes that are isolated in the other networks or that have been separated from the giant component. Of course, from one network to another, the content of the components may vary. Identification is mainly based on hubs and authorities around which they are structured.

To illustrate the evolution of the small components in the different networks, let's consider component 1. It is organized around the hub named `get_PROFESSION_GEOGRAPHICAL-REGION`. It is therefore the starting point of a composition. Its two output parameters (concepts `Profession` and `Geographical-Region`) can interact with input parameters of other operations that either take a unique input parameter (`Profession` or `Geographical-Region`) or two input parameters (`Profession` and `Geographical-Region`). 28 operations are linked to this hub in the exact network as compared to 5 operations in the plugin one. In other words, the plugin relations are less numerous than the exact ones for the two considered concepts. In the fitin network, component number 1 contains 33 operations organized around the hub. Those operations correspond to the 28 operations of the exact network and the 5 operations of the plugin network. Therefore, none of the links simultaneously represent exact and plugin relationships. While this component has a star shape in the exact, plugin and fitin networks, it is composed by two stars in the subsume network. The second star is organized around the `get_RECOMENDEDPRICEINEURO` operation which is an authority (input concept: `Year`, output concept: `RecommendedPriceInEuro`). Bridges between the two stars are made by a set of operations that take inputs from the hub and that provide inputs to the authority. The `get_PROFESSION_TIMEMEASURE` operation for example, takes `Country` as input concept (a sub-concept of `Geographical-Region`) and provides `Profession` and `TimeMeasure` concepts (`TimeMeasure` being a super-concept of `Year`). Subsume relations are more numerous for the considered concepts than the exact and the plugin ones. Indeed, among the 84 operations of this component, 65 are linked to the hub.

Because of the star shape, in small components, compositions are mainly formed with two operations. To increase their efficiency, it is necessary to develop gateways services to aggregate these components to the giant one. Knowledge of the topological structure of both small components and the giant component allow to define the most efficient gateways services for the composition. From the small components side, they must be linked to the hub and authorities.

*5.2.3 Giant component*

Characteristics of the largest components are reported in Table 2. In all the networks, the giant component contains the great majority of links. Note that the proportion of links is computed from the network without isolated nodes. The presence of this large component is a good property. It reflects the ability of a great number of interactions between the operations of the collection, and therefore some useful compositions. We pay a particular attention to the links. They are a key element when it comes to Web services composition. The more links there are, the more possibilities there is to compose Web services. Hence, to be successful when looking for a Web service composition, one expects to have a large number of links in an interaction network.

Whatever the network, the giant component concentrates the vast majority of links as compared to the small components. Indeed, its proportion of links ranges from 95% to 99%. The number of links in the exact and subsume networks is of the same order in the collection while plugin relations are the less numerous. Unsurprisingly, the fitin component contains the largest number of links, due to its definition. Subsume and fitin components are the denser; they are two times more dense than the plugin one. The exact component lies in between.

Table 2. Structure of the giant components of the four operation networks.

| Network | Number of nodes | Number of links | Proportion of links | Density |
|---|---|---|---|---|
| plugin | 369 | 2446 | 99% | 0.0180 |
| exact | 341 | 3426 | 98% | 0.0295 |
| subsume | 329 | 3864 | 95% | 0.0358 |
| fitin | 404 | 5832 | 99% | 0.0358 |

These results provide a new lighting to the problem of the composition search process. Indeed, different strategies can be implemented based either on the quality of the composition or on the search cost. Of course, one can also consider mixed strategies taking into account both criteria. When the quality of composition predominates, one can start the composition search in the exact network, followed by the fitin, then the plugin and finally the subsume network, if goals have not been reached. Indeed, this is the order of relevance for solutions to the compositions queries. If the search cost is the main constraint, it can go from the sparser network to the denser one. In this case, the search process could start in the plugin network, followed by the exact one, if necessary. As fitin and subsume networks have the same density, it is preferable to search in the fitin network because the quality of the composition is higher with the same computational cost.

In order to compute the typical properties of complex networks, we now only consider the giant components. In the following, we may use the word network rather than giant component.

*5.2.4 Small-world*

The four networks have the small-world property. Indeed, they exhibit a small average distance. Table 3 shows the ratio between the networks average distance and the average distance of the corresponding Erdős-Rényi network. The ratio is around one half for the plugin and subsume networks. It is higher for the exact one and reaches almost 1 in the fitin network. Note that the average distance globally increases with the number of links. The exact and plugin links superposition in the fitin network does not result in a reduction of the average distance. We observe quite the opposite. Indeed, the average distance value is almost doubled. The additional links are hence not shortcuts between some remote nodes. They must be plugged in at the periphery of the networks. Overall, the average distance tells us the average minimal number of operations used in order to perform a composition. Roughly speaking, any composition can be obtained with an average of three operations in the plugin, exact and subsume networks and four operations in the fitin one. The diameter values, reported in Table 3, measure the maximum value of the shortest paths between any two nodes of a graph. It informs us about the number of operations required in large compositions. Thus, in the plugin network, all the compositions can be performed using at least 4 operations. 5 operations are needed in the exact and subsume networks and 7 in the fitin network. Note that the diameter values exhibit the same behaviour than the average distance according to the network definition. Indeed the diameter also increases with the number of links. This confirms the growth of the network at its periphery without changing its overall organisation.

*5.2.5 Clustering*

The ratio between the networks clustering coefficient and the clustering coefficient of the Erdős-Rényi networks, reported in Table 3, is always below 1. As Erdős-Rényi networks are not transitive, this clearly demonstrates that all the operation networks are not as well. The fitin component has the highest transitivity, certainly induced by the fact that it has the highest number of links. Nevertheless, the proportion of 3-cliques is negligible; as we can see in Figure 5, nodes are rather organized hierarchically. From the composition process perspective, a low clustering coefficient account for the fact that there is very few situations than a basic composition involving two operations can be performed using one more operation. In this situation, there is very little redundancy which results in a lower robustness against failures. Indeed, if a link is cut in a triangular structure, information can pass through the two other nodes of the triangle.

*5.2.6 Assortativity*

The degree correlation values are reported in Table 3. It is of the same order for the four networks. These negative values indicate that like many real-world networks such as information, technological or biological networks, Web services operation networks are disassortative. Hubs and authorities are preferentially linked to weakly connected nodes rather than being linked together. This behaviour is typical of the one observed in many complex systems emerging from an unplanned organisation. Newcomers tend to aggregate to the structure while favouring elements which possess a strong connectivity. This structure centred on the most influential nodes (hubs and authorities) goes against security. Indeed, the failure of a hub or authority leads to the impossibility of a very large number of compositions.

Table 3. Distance, diameter, clustering and assortativity in the giant components of the four operation networks. Ratio between the distance and the clustering of the components and their counterpart Erdős-Rényi ($X/X_{ER}$).

| Network | Distance | | Diameter | Clustering | | Assortativity |
|---|---|---|---|---|---|---|
| | L | $L/L_{ER}$ | | C | $C/C_{ER}$ | |
| plugin | 1.31 | 0.44 | 3 | 0.018 | 0.48 | -0.48 |
| exact | 1.87 | 0.67 | 4 | 0.022 | 0.36 | -0.43 |
| subsume | 1.38 | 0.56 | 4 | 0.027 | 0.29 | -0.51 |
| fitin | 2.30 | 0.90 | 6 | 0.056 | 0.80 | -0.30 |

*5.2.7 Degree distribution*

Figure 4 (Left) shows the cumulative degree distributions of the operation networks. They all reflect an inhomogeneous behaviour as observed in many real-world networks. Indeed, few nodes have a high degree and the great majority of nodes have a low degree.

Nevertheless, when inspecting the low degree node distribution zone, we observe that a great proportion of median degree nodes stand along with a very low proportion of small degree nodes. This last feature is not usual in real-world networks that exhibit a scale-free degree distribution. To go deeper, we fitted the distributions to a power-law and to an exponential distribution. Figure 4 (Right) shows the exact giant component cumulative degree distribution (blue). The power-law distribution that best fit the empirical data is obtained with an exponent value of 1.1. The best fit for the exponential law exhibit an exponent value of 0.05. We can distinguish two areas delimited by the degree value 10. For degree values below 10, the exponential law is a better fit than the power-law, while for degree values above 10 it is the opposite. Hence, only the tail of the distribution follows a power-law. Note that the degree axis is represented until a value of 100. Indeed, the curves merge from this value. This heavy tail behaviour is typical of real-world networks as compared to random ones. The mixed behaviour for low degree nodes seems to occur because of the re-sampling process. Indeed, the "cloning" of some services has a greater and more visible impact on nodes with few connections. While high degree nodes keep their high degree values, there is a shift for low degree nodes to median values. The three other network degree distributions exhibit the same behaviour.

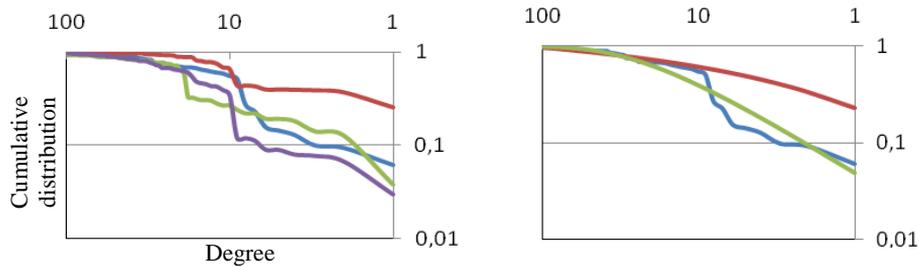

Figure 4. Left: Log-log plot of the cumulative degree distribution in the giant components of operation networks: plugin (red), exact (blue), subsume (green), fitin (purple). Right: Log-log plot of the degree distribution in the giant component of the exact network with power-law fit (red, exponent = 1.1) and exponential fit (green, exponent = 0.05).

Composition search algorithms can be designed to exploit statistical features of the networks structure in order to improve their performances. Such a guided search focuses on a particular fraction of the network chosen because the nodes are likely to rapidly satisfy a goal. In this line, one can take advantage of the highly skewed degree distributions i.e. the presence of highly connected nodes. Indeed, their neighbours account for a significant fraction of all the nodes in the network. This gives opportunities to numerous interactions and as many possibilities to rapidly satisfy a goal. For example, in backward chaining strategies, the authorities can be used as a starting point in the search process while in forward chaining, one will use hubs.

According to the previous studied properties, it appears that the networks are dominated by trees rather than by triangles and that some strongly connected nodes stand aside numerous lightly connected ones. One can observe the phenomenon on Figure 5 where low degree nodes are at the periphery, high degree nodes are concentrated on few spots and median degree nodes are located in between. This heterogeneity has a fundamental role in the propagation phenomena in real-world networks. For example, information or epidemics in social networks, viruses in technological networks like the Internet, can easily and quickly spread out through highly connected nodes. The heterogeneous structure of the complex networks also has an impact on their behaviour that follows a disturbance. These networks are highly resistant to failures (nodes randomly removed) and at the same time extremely fragile to targeted attacks that concentrate on highly connected nodes. In Internet for example, shutdown or dysfunction of local servers can affect the global properties of the communication. The same causes producing the same effects, hubs and authorities play a central role in the composition process and their failure may be critical. Indeed, hubs correspond to operations which can invoke many other operations while authorities correspond to operations that can be invoked by many others. If an operation is a hub, its output is needed by many others. If it becomes unavailable, all these operations cannot be composed anymore, unless other operations providing equivalent parameters exist. Failure on hub operations hence can be very damageable for a composition process. If an operation is an authority, it can be composed with many others. Failures concerning authorities damage all this compositions. The knowledge of those important and sensitive parts of a network is of paramount importance. Their critical position must lead to targeted protection strategies.

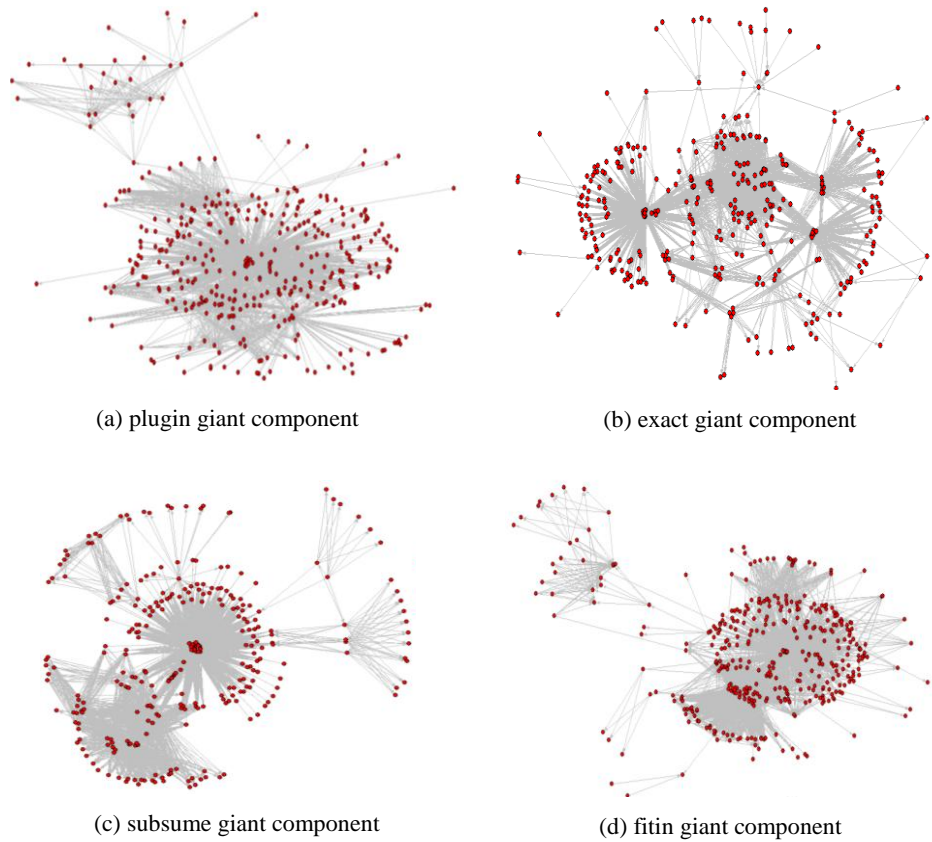

(a) plugin giant component      (b) exact giant component

(c) subsume giant component      (d) fitin giant component

Figure 5. Giant components of operation networks: (a) plugin, (b) exact, (c) subsume, (d) fitin.

*5.2.8 Community structure*

Community detection is performed using Walktrap algorithm (Pons and Latapy, 2005). This dynamic random walk based algorithm, that has been the subject of comparative studies, is known to perform well (Navarro and Cazabet, 2011; Orman, Labatut, and Cherifi, 2011). Table 4 reports the number of communities detected for each network and the associated modularity values. The modularity is high for the exact network, reflecting a well-defined partitioning. The rather low values observed in the other networks are characteristic of weakly cohesive communities. In the fitin network, the number of detected community is comparable to the exact network but with a high decrease of modularity. Walktrap discovered around three times fewer communities in the plugin network and five times less in the subsume network than in the exact one. The higher density observed in the subsume and fitin networks as compared to the exact network has resulted in blurring the community structure. According to the modularity value, we can conclude that the exact network is the only one to exhibit a well-defined community structure. If we look at the community size distribution of the two networks that have a similar number of communities i.e. exact and fitin, we observe that, in both cases, the three largest communities contain more than 80% of the nodes. They stand along with a series of small sized communities. Hence, the distribution appears to be highly inhomogeneous, what is in line to what is generally observed in real-world networks partitioning.

This modular organisation can be exploited throughout the Web services life cycle. Communities of interacting Web services can be substituted to the classical communities based on Web services similarity (Medjahed and Bouguettaya, 2005; Arpinar, Aleman-Meza, and Zhang, 2005; Benatallah, Dumas, and Sheng, 2005; Taher et al., 2006; Bruno et al., 2005; Oldham et al., 2004; Katakis et al., 2009; Konduri and Chan, 2008; Nayak and Lee, 2007; Azmeh et al., 2008), used in the classification process for publication. Rather than grouping Web services belonging to the same domain, these communities gather Web services that preferentially interact. Note that a similar classification has been proposed in (Dekar and Kheddouci, 2008).

Composition search algorithms can also take advantage of the community structure in order to reduce the search space. From this point of view, due to its high modularity value, the exact network is the most appropriate. Important nodes within communities, due to their central position, can be the starting points of a composition search because they share numerous relations with intra-community nodes. Peripheral nodes can play a "smuggler" role at the interface of two communities. The composition search can also be guided by the relation between the semantic content of the communities and the semantic of the requests. Indeed, search performed at the community level first can drastically reduce the search space. To go deeper in this line, we investigated the relationship between identified communities and domains. We observed that globally, those two ways of classifying operations are fairly independent as they do not overlap. The three largest communities of the exact network contain operations from *economy*, *travel* and *education* domains. For medium sized communities, the mixing is more homogeneous. The notion of community, regarding the composition problem, is far more interesting than the notion of domain. A community groups operations that can be composed, while the classification by domains does not induce either an interaction relationship between the operations of a domain or between operations of different domains.

Table 4. Number of communities and modularity of the community structure in the giant components of operation networks.

| Network | Number of communities | Modularity |
|---------|----------------------|------------|
| exact   | 20                   | 0.478      |
| fitin   | 18                   | 0.096      |
| plugin  | 7                    | 0.13       |
| subsume | 4                    | 0.066      |

*5.3 Comparing parameter and operation networks*

The parameter network exhibits the same global structure than the operation networks. Nodes are distributed among a large component, a collection of small components and a set of isolated nodes. Nevertheless, the nodes repartition follows a different distribution. In the following we will concentrate on the exact parameter network and its counterpart, the exact operation network. The proportion of the nodes in these different elements is reported in Table 5.

The parameter network size is equal to the size of the vocabulary used to describe the parameters while the operation network size is the number of operations. This situation leads to a parameter network that is more than two times smaller than the operation network. Many of the 2136 parameters of the collection appear several times. For example, the parameter named `_PRICE` has 130 occurrences and the parameter named `_AUTHOR` has 74 occurrences. As long as they are related to the same concept, they are represented by the same node. The representation of several occurrences of a parameter in one node has a direct consequence on the number of links. Indeed, in some situations, there is less links between parameters than the corresponding number of operations. Figure 6 is an extract of the parameter network with 7 links build with 9 operations. In this example, 3 operations have the parameter named `_COUNTRY` as input and the parameter named `_TIMEMEASURE` as output. In the parameter network they are represented by one link.

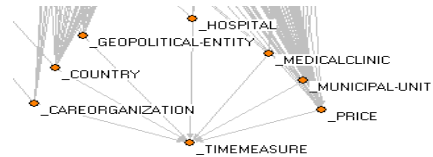

Figure 6. Effect of grouping parameters on the number of links in the parameter network. `_COUNTRY` and `_TIMEMEASURE` nodes respectively represent 2 parameters belonging to 3 distinct operations. The unique link between them represents those 3 operations.

*5.3.1 Isolated nodes*

The proportion of isolated nodes is much lower in the parameter network. It almost contains 12 times less isolated nodes than the operation network. In an operation network, isolated nodes are operations which do not interact, while in a parameter network isolated nodes are related to the message exchange patterns. They belong to operations with only one type of parameter, input or output. For example, the parameter `Dutytax` appears only once in the collection as an output parameter of the `Camerataxedpricedutytax` operation which has no input parameter. Hence, it is represented as an isolated node. As there are few isolated nodes in the network, this indicates that few parameters have those characteristics. The great majority of them are shared by several operations.

*5.3.2 Small components*

The small components in the parameter network contain three times more nodes than in the operation network.

Table 5. Networks size and proportion of nodes in different parts of the exact operation and parameter networks (isolated nodes, small components, giant component).

| Network | Network size | Isolated nodes | Small components | Giant component |
|---|---|---|---|---|
| operation | 785 | 48.79% | 7.77% | 43.44% |
| parameter | 357 | 4.20% | 20.73% | 75.07% |

In Figure 7, small components have been numbered according to their size. We see that they are more numerous in the parameter network. This network contains two times more small components. Furthermore, their size is more homogeneous. The star shape structure is less obvious than the one observed in the operation networks. Nevertheless, we note the presence of a few authorities which reflect two different situations. Indeed, an authority can emerge when different operations share the same output parameter or when a single operation has a lot of input parameters with a single output parameter. Among other differences, smaller components do not emerge from the same domains. For example, the largest component contains three authorities. All its parameters belong to the "unclassified" domain of the collection while in all the operation networks, small components emerge either from the travel or the education domain.

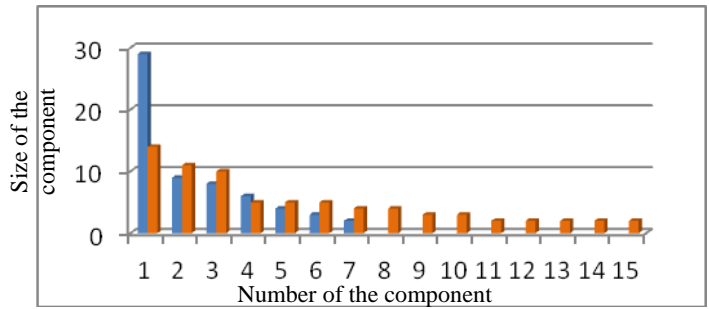

Figure 7. Size distribution of the small components of the exact operation (blue) and parameter (orange) networks.

A big difference is related to the fact that a small component may contain no composition. Indeed, in an operation network, a component necessarily represents one or several compositions. The smallest possible component of two nodes embodies two operations in an interaction relation. This is not the case in the parameter network where a component may represent a single operation. If it contains several operations, they share some parameters, but this does not imply that a composition emerges from it. This case is illustrated by the small component given in Figure 8.

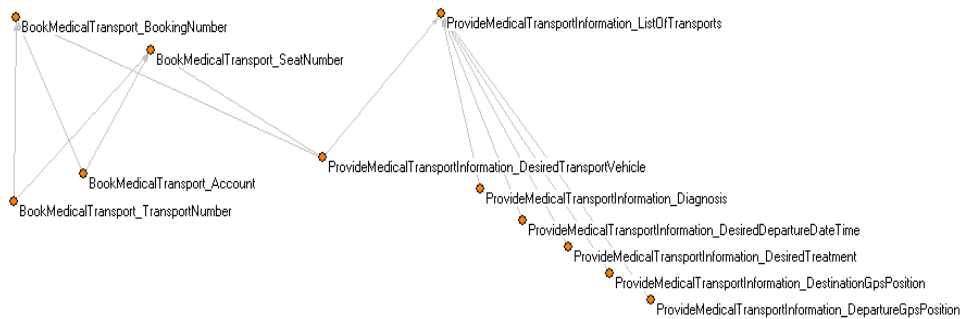

Figure 8. A small component of 2 operations in the parameter network with no interaction relationship. Left side: `Bookmedicaltransport` operation with 5 parameters (3 inputs (down) and 2 outputs (up)). Right side: `Providemedicaltransportinformation` operation with 7 parameters (6 inputs (down) and 1 output (up)). Middle: 1 input parameter (`ProvideMedicalTransportInformation_DesiredTransportVehicle`) shared by the two operations.

This 11 nodes component contains two operations (`Bookmedicaltransport` and `Providemedicaltransportinformation`). `Bookmedicaltransport` has three inputs (`BookMedicalTransport_TransportNumber,BookMedicalTransport_Account,ProvideMedicalTransport_desiredTransportVehicle`) and two outputs (`BookMedicaltransport_bookingNumber, BookMedicalTransport_Seatnumber`). `Providemedicaltransportinformation` has one output

(`provideMedicalTransportation_ListOfTransports`) and 6 inputs. Both operations share a single input parameter (`ProvideMedicalTransport_desiredTransportVehicle`). As none of the parameters are input of one operation and output of the other, it is not possible to compose these two operations. Small components do not necessarily contain operations implied within a composition relationship in a parameter network while it is always the case for an operation network. One must keep in mind this essential difference for composition search.

*5.3.3 Giant component*

The characteristics of the giant component of exact parameter and operation networks are reported in Table 6. Note that the number of links is almost six times smaller in the parameter network than in the operation network. This results in a sparser network. This is due to the fact that parameters of different operations are grouped into the same nodes and consequently links represent several operations. Although the proportion of links of the giant component is 10% higher for the operation network, in both cases the giant component concentrates the vast majority of links. Those features can be observed on the representation of the two giant components in Figure 9. The parameter network is smaller with less nodes and links. This can be considered as advantageous for the composition process. Indeed, searching for a composition in a smaller network is easier. However, links do not have the same meaning in both networks. In an operation network, a link account for the fact that an operation provides all the required parameters to invoke another one, while in a parameter network, a link just relates that there is an input/output relationship between two parameters. Hence, at least two links are needed to represent a composition in a parameter network. Furthermore, one needs to maintain the information about the operations that are represented by a link.

Table 6. Structure of the giant components of the exact operation and parameter networks: number of nodes, number and proportion of links, density.

| Network | Number of nodes | Number of links | Proportion of links | Density |
|---|---|---|---|---|
| operation | 341 | 3426 | 98% | 0.0295 |
| parameter | 268 | 621 | 88% | 0.0086 |

*5.3.4 Small-world*

Like the operation network, the parameter network exhibits the small-world property. As shown in Table 7, the ratio between the average distance of the giant component and the average distance of the corresponding Erdős-Réyni network is far below 1. Despite that it is sparser, the average distance of the parameter network is just slightly higher as compared to the operation network. Such results suggest that many shortcuts exist to join efficiently the different areas of the network. Hence, one can produce some parameters of interest using a relatively small number of operations. Nevertheless these results should be treated with caution. Indeed, generally, the operations do not possess a single input and a single output. It is therefore difficult to extrapolate a statistic on the number of operations involved in composition from the average distance.

*5.3.5 Clustering*

The clustering coefficients of exact parameter and operation networks are reported in Table 7. They are very low and therefore the networks are not transitive. In the parameter network, the coefficient is slightly higher and the ratio between the coefficient of the network and the one of the Erdős-Réyni network is above 1. Nevertheless, this does not mean that there is a great proportion of triangles. As confirmed by a networks visualisation in Figure 9, nodes are rather hierarchically organized.

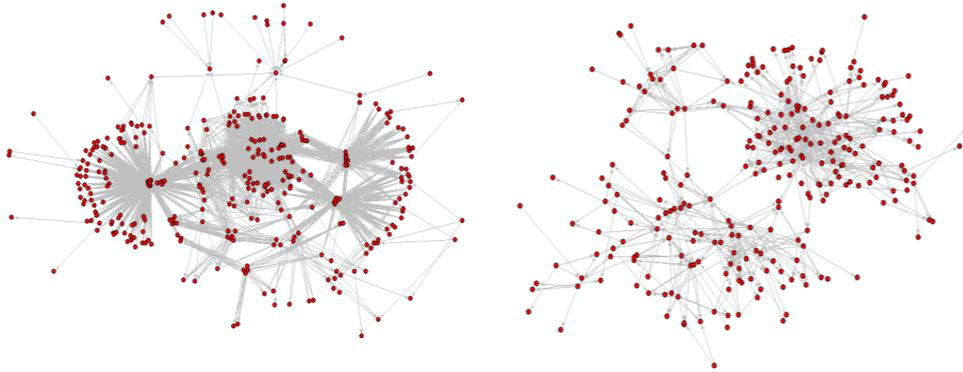

Figure 9. Exact operation (left) and parameter (right) giant components.

*5.3.6 Assortativity*

The negative degree correlation values reported in Table 7 reveal a disassortative behaviour in the parameter network, like in the operation network. Nodes tend to connect with other nodes having different degree values. However, this is far less pronounced for the parameter network.

Table 7. Distance, clustering and assortativity in the giant components of the operation and parameter networks. Ratio between the distance and the clustering of the components and their counterpart Erdős-Réyni (ER).

| Network | Distance | | Clustering | | Assortativity |
|---|---|---|---|---|---|
| | L | L/L$_{ER}$ | C | C/C$_{ER}$ | |
| operation | 1.87 | 0.67 | 0.022 | 0.36 | -0.43 |
| parameter | 1.97 | 0.31 | 0.031 | 1.55 | -0.22 |

*5.3.7 Degree distribution*

The degree distribution of exact parameter and operation networks is non-homogeneous with heavy tail behaviour. However, the parameter network has the scale-free property. Its degree distribution follows a power-law. The maximum likelihood estimate of the power-law coefficient value is $\gamma = 3.04$. The p-value of the Kolmogorov-Smirnov test (0.84) shows that it is a good fit to the empirical data. Figure 10 presents the plots of the empirical degree distribution and the estimated power-law in a log-log scale. In such a representation, the signature of a power-law is a straight line. We think that the impact of the re-sampling process of the collection on the degree distribution is less visible on the parameter network than on the operation networks. Indeed, when a Web service is duplicated, there is no impact on the parameter network while there will be a new node and also new links in the operation network. This phenomenon necessarily affects the degree distribution of the operation network, while the parameter network is insensitive to this modification.

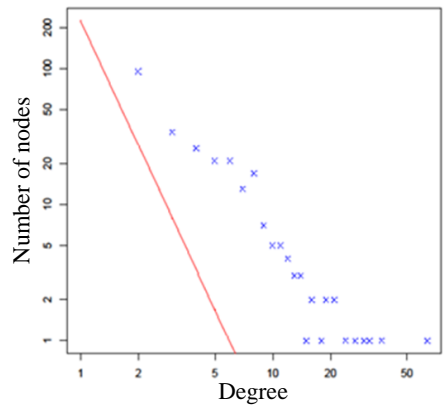

Figure 10. Log-log plot of the degree distribution in the giant component of the parameter network (cross) and estimated power-law with exponent value 3.04 (line).

In the operation network, the strongly connected nodes (hubs and authorities) represent operations than can participate to several compositions. In the parameter network, strongly connected nodes represent parameters used by many operations either as input or as output. Both networks are dominated by trees with the presence of hubs and authorities. In the parameter network, hubs correspond to parameters used as input by many operations while authorities correspond to parameters being outputs of many operations. `Country` and `Price` are such remarkable parameters. A hub is a parameter that corresponds to different situations. It is an input of an operation producing several output parameters, or it is an input of several operations producing one or several output parameters. The production of many other parameters depends on its presence. If it becomes unavailable, all these parameters cannot be produced anymore, unless other operations produce it. The failure of operations producing hub parameters can be very detrimental. An authority is a parameter that corresponds to the following situations. It is an output of an operation taking several input parameters, or it is an output of several operations taking one or several input parameters. If a parameter is an authority, its production depends on many others or there are many operations able to produce it. Hence, depending on the situation, failures on operations producing authorities can have less consequence.

*5.3.8 Community structure*

As shown in Table 8, the number of communities is lower in the parameter network. Nevertheless, it remains of the same order than in the operation network. The modularity for both networks falls in the range of [0.3-0.7]. It is higher in the parameter network. In other words, detected communities are more cohesive. Figure 11 represents the community size in the exact parameter and operation networks. Communities have been ranked and numbered according to their size in ascending order (biggest community numbered 1). The community size distribution in the parameter network follows the same trend than in the operation network. It is quite inhomogeneous with few highly populated communities and more numerous being slightly populated. It is globally lower in the parameter network for the biggest communities (Cherifi and Santucci, 2013). Considering a composition search, the smallest number of communities and the higher modularity favours the parameter network. Indeed, the space to be explored is smaller and better defined. Regarding the domains, communities in the parameter network are more domains-centred. Indeed, the network is organized around a common vocabulary, i.e. the parameters, which is specific to each domain.

Table 8. Number of communities and modularity in the giant components of exact networks of parameters and of operations.

| Network | Number of communities | Modularity |
|---|---|---|
| operation | 20 | 0.478 |
| parameter | 16 | 0.618 |

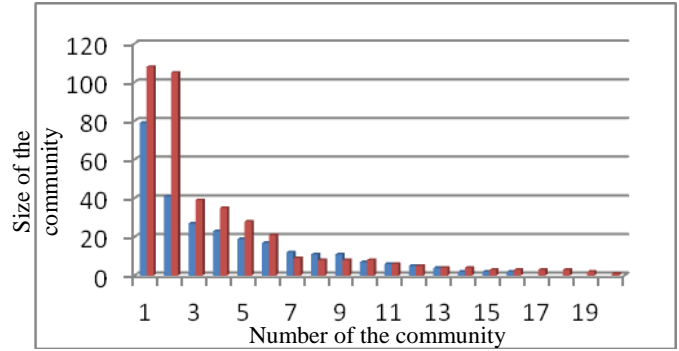

Figure 11. Communities size distribution for the 20 communities of the exact operation network (purple) and for the 16 communities of the exact parameter network (blue).

We summarize the main characteristics of the studied networks that may affect a composition search process and that are likely to discriminate them. Two main features are in favour of a parameter network. Its smaller size and its lower density can allow a faster process because the search space is reduced. In the case of a composition search guided by the community structure, the higher modularity facilitates the identification of communities. Nevertheless, one needs additional information about the links and which operations they represent. In an operation network, those advantages are compensated by the fact that one needs to explore only one link to find a basic composition of two operations while two links are necessary in a parameter network. This implies that in an operation network, all the small components contain compositions and could eventually be considered. In a parameter network, one must keep in mind that some of them are not usable because no interaction relationship emerges. Finally, one can take advantage of the high node centrality in the two types of networks. Indeed, possible interactions are more numerous from nodes with high degree.

## 6. Conclusion

In this paper, we report on experiments on local and global properties of semantic Web services interaction networks using a benchmark of real-world Web services descriptions. Topological properties of the parameter network where the nodes are input or output parameters of Web services and operation networks where the nodes are the Web service operations, have been investigated. This comparative study shows that all the interaction networks share the same global structure. A giant component stands along with some small components and a set of isolated nodes. Our study indicates that all the giant components share the same characteristics of most real-world complex networks: the small-world property that accounts for a short average distance between any two nodes of the network and an inhomogeneous distribution on the links between the nodes, with a few hubs which are highly connected. Furthermore they are more or less organised into communities. Their low clustering coefficient values are similar to the ones obtained for technological networks such as the Internet. The networks are disasortatives; in other words, similar nodes (i.e. with the same number of links) do not tend to associate together.

Based on these topological features, we give some directions about the use of those different network structures for the composition search process. It can take advantage of the presence of highly connected nodes observed in all the networks. Because of their numerous neighbours, they give opportunity to rapidly reach a given goal, if used as a starting point of the search process. The community structure can also be a guide for the composition search. Indeed, search can start at the community level. Such a strategy can drastically reduce the search space.

Operation and parameter networks offer different views of the composition process. A parameter network represents all the possible interactions that can occur between the set of operations while an operation network represents the interactions between all the instances of operations. The former carries the information about the existence of a composition while the latter tells us which operations are involved in this composition. Both can be conjointly involved in a composition search. Indeed, the smaller size of the parameter network makes it suitable to be mined first in order to know if there is a solution for the composition request. If so, candidate Web services can be extracted from an operation network.

Due to its smaller size and its lower density, search in the parameter network is more efficient than in an operation network. One must nevertheless bear in mind that additional information on links is needed to know which operations they represent. Note that, among other benefits, it exhibits the most cohesive communities.

The comparison of the four operation networks turns at the advantage of the exact network. It contains the best composition solutions and exhibits the most favourable structure for a composition search. The other operation networks can be considered, if needed, even if the compositions are less effective.

The network structure investigation also revealed the presence of hubs and authorities. These are the cornerstones regarding dynamic processes like failures and attacks that occur in networks. Although providers may be interested by the development of such highly used operations, they are in turn sensitive parts of a network because of their high connectivity. Hence, they must be identified and targeted protection strategies may be specifically developed.

The main contributions of this work is to provide a thorough study of parameter and operation networks structure by highlighting topological differences, and to give guidelines on the use of those features to guide a composition search process. We are extending this work in two directions. In the first one we are using the networks topology knowledge for composition search. Indeed, based on networks characteristics, different strategies can be devised in order to explore it more efficiently. For example, the community structure can be exploited in order to reduce the search space. Furthermore, hubs can be good starting points for the exploration. The second one is about generating semantic Web service descriptions to produce benchmarks that meet network properties and whose purpose is to test search composition algorithms.

**References**

Albert, R., Jeong, H. & Barabasi, A.-L., 1999. The diameter of the world wide web. *Nature*, 401(September), pp.130–131.

Arpinar, I., Aleman-Meza, B. & Zhang, R., 2005. Ontology-driven web services composition platform. *Inf. Syst. E-Business Management*, vol. 3.

Azmeh, Z. et al., 2008. WSPAB: A Tool for Automatic Classification & Selection of Web Services Using Formal Concept Analysis. In *Sixth European Conference on Web Services*. IEEE, pp. 31–40.

Benatallah, B., Dumas, M. & Sheng, Q.Z., 2005. Facilitating the Rapid Development and Scalable Orchestration of Composite Web Services. *Distributed and Parallel Databases*, 17(1), pp.5–37.

Boccaletti, S. et al., 2006. Complex networks: Structure and dynamics. *Physics Reports*, 424(4-5), pp.175–308.

Bruno, M. et al., 2005. An Approach to support Web Service Classification and Annotation. In *2005 IEEE International Conference on e-Technology, e-Commerce and e-Service*. IEEE, pp. 138–143.

Cherifi, C., Labatut, V. & Santucci, J.-F., 2010a. Benefits of Semantics on Web Service Composition from a Complex Network Perspective. In F. Zavoral, J. Yaghob, & E. El-Qawasmeh, eds. *International Conference on Networked Digital Technologies*. Prague: Springer, pp. 80–90.

Cherifi, C., Labatut, V. & Santucci, J.-F., 2010b. Web Services Dependency Networks Analysis. In M. G. Nalbant & T. Kara, eds. *International Conference of New Media and Interactivity*. Istanbul : Marmara University, pp. 115–120.

Cherifi, C., Rivierre, Y. & Santucci, J.-F., 2011. WS-NEXT, a Web Services Network Extractor Toolkit. In *International Conference on Information Technology*. Amman.

Cherifi, C. & Santucci, J.-F., 2013. Community Structure in Interaction Web Service Networks. *To be appeared in Int. J. of Web Based Communities*, 9(3).

Costa, L. da F. et al., 2007. Characterization of complex networks: A survey of measurements. *Advances in Physics*, 56(1), pp.167–242.

Couto, F.M. & Silva, M.J., 2011. Disjunctive shared information between ontology concepts: application to Gene Ontology. *Journal of biomedical semantics*, 2(1), p.5.

Dekar, L. & Kheddouci, H., 2008. A Graph b-Coloring Based Method for Composition-Oriented Web Services Classification. In A. An et al., eds. Springer Berlin Heidelberg, pp. 599–604.

Fan, J. & Kambhampati, S., 2005. A snapshot of public web services. *ACM SIGMOD Record*, 34(1), p.24.

Farrell, J. & Lausen, H., 2007. *Semantic Annotations for WSDL and XML Schema*, Available at: http://www.w3.org/TR/sawsdl/.

Fortunato, S., 2010. Community detection in graphs. *Physics Reports 486*, pp.75–174.

Gekas, J. & Fasli, M., 2007. Employing Graph Network Analysis for Web Service Composition. *International Journal of Information Technology and Web Engineering*, 2(4 ), p.20.

Guimerà, R. et al., 2003. Self-similar community structure in a network of human interactions. *Physical Review E*, 68(6).

Hashemian, S.V. & Mavaddat, F., 2005. A Graph-Based Approach to Web Services Composition. In *2005 Symposium on Applications and the Internet*. IEEE, pp. 183–189.

Hau, J. & Lee, W., 2005. A Semantic Similarity Measure for Semantic Web Services. In *Web Service Semantics Workshop at WWW*.

Hess, A., Johnston, E. & Kushmerick, N., 2004. ASSAM: A tool for semi-automatically annotating semantic web services. In S. A. McIlraith, D. Plexousakis, & F. van Harmelen, eds. *International Semantic Web Conference*. Hiroshima: Springer.

IEEE International Conference on e-Business Engineering, 2005. ICEBE'05. (2005). Available at: http://ieeexplore.ieee.org/xpl/mostRecentIssue.jsp?punumber=10403.

InfoEther & Technologies, B., 2004. SemWebCentral. Available at: http://wwwprojects.semwebcentral.org/.

Katakis, I. et al., 2009. On the Combination of Textual and Semantic Descriptions for Automated Semantic Web Service Classification. In *Artificial Intelligence Applications and Innovations*. Boston: Springer, pp. 95–104.

Kil, H. et al., 2009. Graph Theoretic Topological Analysis of Web Service Networks. *World Wide Web*, 12(3), pp.321–343.

Konduri, A. & Chan, C., 2008. Clustering of Web Services Based on WordNet Semantic Similarity, Akron: University of Akron, USA.

Kwon, J. et al., 2007. PSR : Pre-computing Solutions in RDBMS for Fast Web Services Composition Search. In *International Conference on Web Services* . Salt Lake City, Utah, USA, pp. 808–815.

Küster, U., König-Ries, B. & Krug, A., 2008. OPOSSum - An Online Portal to Collect and Share SWS Descriptions. In *International Conference on Semantic Computing*. Santa Clara, California, USA: IEEE, pp. 480–481.

Lausen, H., Polleres, A. & Roman, D., 2005. Web Service Modeling Ontology (WSMO). Available at: http://www.w3.org/Submission/WSMO/.

Liu, J. & Chao, L., 2007. Design and Implementation of an Extended UDDI Registration Center for Web Service Graph. In *International Conference on Web Services*. Salt Lake City, Utah, USA: IEEE, pp. 1174–1175.

Martin, D. et al., 2004. *OWL-S: Semantic Markup for Web Services*, Available at: http://www.w3.org/Submission/OWL-S/.

Medjahed, B. & Bouguettaya, A., 2005. A Dynamic Foundational Architecture for Semantic Web Services. *Distributed and Parallel Databases*, 17(2), pp.179–206.

Navarro, E. & Cazabet, R., 2011. Détection de communautés, étude comparative sur graphes réels. *Information interaction intelligence*, 11(1), pp.77–93.

Nayak, R. & Lee, B., 2007. Web Service Discovery with additional Semantics and Clustering. In *IEEE/WIC/ACM International Conference on Web Intelligence*. Silicon Valley, USA: IEEE, pp. 555–558.

Newman, M., Barabási, A.-L. & Watts, D.J., 2011. *The Structure and Dynamics of Networks*, Princeton University Press.

Newman, M.E.J., 2004. Detecting community structure in networks. *The European Physical Journal B Condensed Matter*, 38(2), pp.321–330.

Newman, M.E.J., 2006. Modularity and community structure in networks. *Proceedings of the National Academy of Sciences of the United States of America*, 103(23), pp.8577–8582.

Newman, M.E.J., 2003. The Structure and Function of Complex Networks. *SIAM Review*, 45(2), p.167.

Oldham, N. et al., 2005. METEOR-S Web Service Annotation Framework with Machine Learning Classification. In *Lecture Notes in Computer Science : Semantic Web Services and Web Process Composition* . pp. 137–146.

Orman, G., Labatut, Vincent & Cherifi, H., 2011. Qualitative Comparison of Community Detection Algorithms. In H. Cherifi, J. M. Zain, & Eyas El-Qawasmeh, eds. *Digital Information and Communication Technology and Its Applications*. Springer, pp. 265–279.

Paolucci, M. et al., 2002. Semantic Matching of Web Services Capabilities. In *The Semantic Web - ISWC 2002 - LNCS*. Springer, pp. 333–347.

Pons, P. & Latapy, M., 2005. Computing communities in large networks using random walks. *Journal of Graph Algorithms and Applications*, 10(2), pp.191–218.

Rada, R. et al., 1989. Development and application of a metric on semantic nets. *IEEE Transactions on Systems, Man, and Cybernetics*, 19(1), pp.17–30.

Resnik, P., 1995. Using Information Content to Evaluate Semantic Similarity in a Taxonomy. In *Proceedings of the 14th International Joint Conference on Artificial Intelligence*. pp. 448–453.

Scott, J., 2000. *Social Network Analysis: a Handbook.* SAGE.

Shvaiko, P. & Euzénat, J., 2005. A Survey of Schema-Based Matching Approaches S. Spaccapietra, ed. *Journal on Data Semantics*, IV, pp.146–171.

Taher, Y. et al., 2006. Towards an Approach for Web Services Substitution. In *10th International Database Engineering and Applications Symposium*. IEEE, pp. 166–173.

Talantikite, H., Aissani, D. & Boudjlida, N., 2009. Semantic annotations for web services discovery and composition. *Computer Standards Interfaces*, 31(6), pp.1108–1117.

Watts, D.J. & Strogatz, S.H., 1998. Collective dynamics of "small-world" networks. *Nature*, 393(6684), pp.440–2.